# Macroscopic visualization of fast electrochemical reaction of SrCoO$_x$ oxygen sponge

Qian Yang[1], Hai Jun Cho[1,2], Hyoungjeen Jeen[3*], and Hiromichi Ohta[1,2*]

[1]Graduate School of Information Science and Technology, Hokkaido University, N14W9, Kita, Sapporo 060−0814, Japan

[2]Research Institute for Electronic Science, Hokkaido University, N20W10, Kita, Sapporo 001−0020, Japan

[3]Department of Physics, Pusan National University, Busan, 46241, Korea

**Strontium cobaltite (SrCoO$_x$) is known as a material showing fast topotactic electrochemical Redox reaction so-called 'oxygen sponge'. Although atomic scale phenomenon of the oxidation of SrCoO$_{2.5}$ into SrCoO$_3$ is known, the macroscopic phenomenon has not been clarified yet thus far. Here, we visualize the electrochemical oxidation of SrCoO$_x$ macroscopically. SrCoO$_x$ epitaxial films with various oxidation states were prepared by the electrochemical oxidation of SrCoO$_{2.5}$ film into SrCoO$_{3-\delta}$ film. Steep decrease of both resistivity and the absolute value of thermopower of electrochemically oxidized SrCoO$_x$ epitaxial films indicated the columnar oxidation firstly occurred along with the surface normal and then spread in the perpendicular to the normal. Further, we directly visualized the phenomena using the conductive AFM. This macroscopic image of the electrochemical oxidation would be useful to develop a functional device utilizing the electrochemical redox reaction of SrCoO$_x$.**

Transition metal oxides (TMOs)[1] with electrochemically controllable physical properties have been attracted attention as active materials for the electrochemical memory devices. Among many TMOs, strontium cobaltite (SrCoO$_x$)[2, 3] shows three kinds of optical, electrical, and magnetic phases when oxidized/protonated; SrCoO$_{2.5}$ is brown-colored electrical insulator and it shows antiferromagnetic behaviour, oxidized SrCoO$_{3-\delta}$ is black-colored metal and it shows ferromagnetic behaviour, and protonated HSrCoO$_{2.5}$ is almost transparent electrical insulator and it shows weak ferromagnetic behavior.[4, 5] As schematically shown in **Fig. 1(a)**, there are oxygen vacancies ($V_{O2-}$) in SrCoO$_{2.5}$ crystal.[6] The vacancies are occupied with oxygen after the oxidation (SrCoO$_{3-\delta}$)[3]. Valence states of cobalt ion change from 3+ into 4+ by the oxidation of SrCoO$_{2.5}$ into SrCoO$_{3-\delta}$. It is known that the topotactic redox reaction of SrCoO$_{2.5}$ ↔ SrCoO$_{3-\delta}$ under controlled atmosphere occur at relatively low temperature (~200 °C), thus, SrCoO$_x$ is called 'oxygen sponge'.[4] Since this redox reaction of SrCoO$_x$ can be controlled an electrochemical way as well at room temperature,[3, 7, 8] several unique electromagnetic memory devices have been proposed.[5, 9, 10]

In 2016, Katase *et al.*[9] demonstrated an SrCoO$_x$ thin film based memory device, which can store both electrical conductivity and magnetic properties simultaneously. They fabricated a three terminal thin-film transistor structure of the SrCoO$_{2.5}$ epitaxial films grown on (001) SrTiO$_3$ using a porous oxide film as the gate insulator. Since the pores in the porous oxide film was filled with water,[11] the following electrochemical oxidation reaction occurred during the negative voltage was applied to the gate electrode.

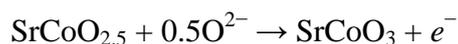

The fully oxidized SrCoO$_3$ showed metallic electron conduction and ferromagnetic behaviour below room temperature. And the proposed electrochemical memory device showed fast (~2 – 3 s) redox reaction under applying relatively low voltage of ±3 V.[9]

In order to further improve the device characteristics and to develop practical devices, the electrochemical oxidation needs to be visualized macroscopically. Usually transmission electron microscopy (TEM) observation is powerful tool to visualize the electrochemical redox reaction of a material.[12, 13] However, a TEM cannot be used for $SrCoO_{3-\delta}$ because $SrCoO_{3-\delta}$ easily release oxygen and becomes $SrCoO_{2.5}$ phase when electron beam is irradiated. On the other hand, conductive atomic force microscopy (AFM) is also a powerful tool to analyze the conductive region in a material.[14] Although the film surface of $SrCoO_x$ needs to be clean to be observed by conductive AFM, the film surface is usually covered with a solid[9] or liquid[3, 5, 15] electrolyte. In order to visualize the electrochemical oxidation macroscopically, we analyzed thermopower of $SrCoO_x$ film with various oxidation states because thermopower is sensitive to the volume of conducting material in an insulating material/conducting material mixture.[16-21] Further, we used an epitaxial film of $SrCoO_{2.5}$ grown on single crystalline solid-electrolyte to modulate the oxidation states of $SrCoO_x$. In order to keep the $SrCoO_x$ film surface clean and to measure conductive AFM, we did not deposit any metal electrode on the surface.

Here we investigate the electrochemical oxidation reaction of $SrCoO_x$ macroscopically by the thermopower analysis of $SrCoO_x$ films with various oxidation states. As the result, we successfully visualized the electrochemical oxidation reaction of $SrCoO_x$; The columnar oxidation firstly occurred along with the surface normal and then spread in the perpendicular to the normal direction. Further, we directly visualized the phenomena using the conductive AFM. This macroscopic image of the electrochemical oxidation would be useful to develop a functional device utilizing the electrochemical redox reaction of $SrCoO_x$.

$SrCoO_x$ epitaxial films with various oxidation states were prepared by the electrochemical oxidation of ~40-nm-thick $SrCoO_{2.5}$ films, which were heteroepitaxially grown on (001) oriented Yttria−Stabilized Zirconia (YSZ) single crystal substrates with 6-nm-thick 8 mol% Gd-doped $CeO_2$ (GDC) by pulsed laser deposition technique.[22-24] In order to use the YSZ substrate as the oxide ion conducting electrolyte, we deposited 55-nm-thick porous Pt film (relative density ~53%) with comb shaped pattern was deposited on the back side of the YSZ substrate by the DC sputtering at room temperature. Then, we put the sample on an Al foil and heated the sample at 300 °C in air as shown in the inset of **Fig. 1b**. It should be noted that the $SrCoO_x$ film surface kept clean after the electrochemical oxidation. Negative voltage (−5 V or −10 V) was applied to the $SrCoO_{2.5}$/GDC/YSZ multilayer structure using the porous Pt electrode and Al foil electrode. The current density was up to −20 μA $cm^{-2}$. The electron density $Q$ was calculated as

$$Q = \frac{I \cdot t}{e \cdot V}$$

where $I$ is the flown current, $t$ is applied time, $e$ is the electron charge, and $V$ is the volume of the film (1 cm × 1 cm × ~40 nm). We modulated the oxidation state by changing the $Q$ as $7.20 \times 10^{20}$ $cm^{-3}$ for sample A, $1.51 \times 10^{21}$ $cm^{-3}$ for sample B, $3.60 \times 10^{21}$ $cm^{-3}$ for sample C, $6.95 \times 10^{21}$ $cm^{-3}$ for sample D, and $2.06 \times 10^{22}$ $cm^{-3}$ for sample E, respectively, as shown in **Fig. 1b**. After the oxidation treatment, we quenched the sample to room temperature.

In order to clarify the crystallographic information of the resultant $SrCoO_x$ films with various oxidation states, we performed high-resolution X-ray diffraction (XRD) measurements (**Fig. 2**). Only intense diffraction peaks of 00*l* Brownmillerite (BM)-$SrCoO_{2.5}$ or 00*l* Perovskite (PV)-$SrCoO_{3-\delta}$ are seen in the out-of-plane XRD patterns (**Fig. 2a**) together with 00*l* GDC / 00*l* YSZ in all cases, indicating strong *c*-axis orientation of the $SrCoO_x$ films. The as grown sample was single phase $SrCoO_{2.5}$ and the sample E was single phase $SrCoO_{3-\delta}$. Samples A, B, C, and D were the mixture composed of BM-$SrCoO_{2.5}$ and PV-$SrCoO_{3-\delta}$. The 200 diffraction peak of $SrCoO_{2.5}$ ($d$ = 1.9305 Å, 1.1 % shorter than that of bulk) was separately observed with 220 diffraction peak of YSZ ($d$ = 1.8248 Å) in the in-plane XRD patterns (data not shown), indicating that the $SrCoO_{2.5}$ film was grown on the GDC/YSZ substrate with 1.1 % compressive strain. **Figure 2b** is the magnified XRD patterns around $008_{BM}$ or $002_{PV}$ Bragg peak. With increasing the oxidation state from as grown to sample E, the peak intensity of $008_{BM}$ decreased whereas that of $002_{PV}$ increased instead. The $d$-spacing of $008_{BM}$ was 0.1973 nm and that of $002_{PV}$ was 0.1910 nm, ~3.2 % shorter than $008_{BM}$ (see **Supplementary Fig. S1a** and **S1b**). From the $d$-spacing of $002_{PV}$, the lattice constant of the oxidized film is 0.382 nm, which is closed to the lattice constant of an $SrCoO_3$ single crystal (0.3829 nm).[25] It should be noted that average crystal tilting ($\equiv$ full width at half maximum of the X-ray rocking curve, see **Supplementary Fig. S1c** and **1d**) of both $SrCoO_{2.5}$ and $SrCoO_{3-\delta}$ films is ~0.7 °, which does not show any difference between the as-grown and oxidized.

Then, we performed the X-ray reflectivity measurements of as grown $SrCoO_{2.5}$/GDC and oxidized $SrCoO_{3-\delta}$/GDC (see **Supplementary Fig. S2**) to analyze the bilayer structure. The thickness of the $SrCoO_{2.5}$ film was 43.9 nm whereas that of the $SrCoO_{3-\delta}$ was 42.5 nm. The film thickness decreased ~3.2 % due to decreasing of the $d$-spacing, indicating that the $SrCoO_{2.5}$ was fully oxidized into $SrCoO_{3-\delta}$. Thickness of the GDC film was 6 nm, which did not change after the oxidation. We also analyzed the surface morphology of the films using AFM. Several hundred nanometer sized grain structure were seen in the topographic AFM images of both $SrCoO_{2.5}$ and $SrCoO_{3-\delta}$ films (see **Supplementary Fig. 3a** and **3b**). The surface morphologies are basically same. Bright streaks in the reflection high energy electron diffraction (RHEED) pattern indicate that the $SrCoO_{2.5}$ was heteroepitaxially grown on the substrate with smooth surface (inset of **Supplementary Fig. S3b**). These results indicate that topotactic oxidation occurred electrochemically without any crystallographic damage.

In order to further analyze the crystallographic information of the SrCoO$_x$ films, we performed reciprocal space mapping (RSM) measurements. **Figures 2c − 2h** are the RSM of (c) as grown, (d) sample A, (e) sample B, (f) sample C, (g) sample D, and (h) sample F. The RSMs showed that the SrCoO$_x$ films were heteroepitaxially grown on the GDC/YSZ with the epitaxial relationship of (001)[100] SrCoO$_{2.5}$ ∥ (001)[110] GDC/YSZ. In **Figs. 2c** and **2d**, only intense diffraction spot of 11$\underline{12}_{BM}$ is observed. On the other hand, only intense diffraction spot of 103$_{PV}$ is observed in **Fig. 2h**. In **Figs. 2f – 2g**, both diffraction spots are seen, clearly indicating that these films are the mixture composed of BM-SrCoO$_{2.5}$ and PV-SrCoO$_{3-\delta}$ as concluded above.

Since the as grown sample was single phase SrCoO$_{2.5}$ and the sample E was single phase SrCoO$_{3-\delta}$, we calculated the conversion rate from BM-SrCoO$_{2.5}$ into PV-SrCoO$_{3-\delta}$ using the ratio of the diffraction peak area of 008 BM-SrCoO$_{2.5}$ and 002 PV-SrCoO$_{3-\delta}$ as a function of the quantity of electricity over required electricity (see **Supplementary Fig. S4a**). With increasing the electricity, the peak area of 002$_{PV}$ increases rapidly and saturates whereas that of 008$_{BM}$ decreases rapidly and approaches to zero. The conversion rate was calculated as (peak area of 002$_{PV}$)/(peak area of 002$_{PV}$ for sample E) as shown in **Supplementary Fig. S4b**. The conversion rate of as grown was 0 %, sample A was 24.4 %, sample B was 34.5 %, sample C was 60.0 %, sample D was 84.7 %, and sample E was 100 %.

From the XRD results, we extracted several size information including lattice constant in both in-plane and out-of-plane directions and grain size. **Figure 3** summarizes the changes in (a) scattering vector in the in-plane direction ($q_x/2\pi$), (b) scattering vector in the out-of-plane direction ($q_z/2\pi$) (11$\underline{12}_{BM}$ or 103$_{PV}$), (c) *d*-spacing value and (d) inverse of the integral width (008$_{BM}$ or 002$_{PV}$). The inverse of the integration width is equal to the coherence length in the out-of-plane direction of the film (Scherrer's equation). The $q_x/2\pi$ (**Fig. 3a**) and $q_z/2\pi$ (**Fig.**

3b) values of as grown and samples B − E were similar to those of the bulk values. However, it should be noted that the $q_x/2\pi$ of $11\underline{12}_{BM}$ for the sample A (24.4 %) is small as compared to the other samples. Although the out-of-plane lattice parameter was similar to the bulk, the in-plane lattice parameter of sample A is 0.4052 nm, which is ~3.8 % larger than that of bulk BM-SrCoO$_{2.5}$ (0.3905 nm). The out-of-plane lattice parameter of both BM-SrCoO$_{2.5}$ and PV-SrCoO$_{3-\delta}$ gradually decreased with increasing the conversion rate (**Fig. 3c**).

**Figure 3d** shows the inverse of the integration width of 008$_{BM}$ or 002$_{PV}$ Bragg diffraction peak. If the electrochemical oxidation occurs two-dimensionally, the systematic change in the integration width can be observed. The values at 0 % (~19 nm) and 100 % (~17 nm) are smaller than the film thickness (~37 nm), which was calculated from the X-ray reflectivity measurements. This is probably due to the Bragg diffraction peak contains the component of lattice distortion. The (integration width)$^{-1}$ of 008$_{BM}$ did not change until the conversion rate reached 40 %. After that (integration width)$^{-1}$ of 008$_{BM}$ gradually decreased with increasing conversion rate. On the other hand, (integration width)$^{-1}$ of 002$_{PV}$ was ~4 nm when 24.4 %. But it increased above 10 nm with increasing conversion rate. And finally it reached 17 nm. Even if qualitative increase in (integration width)$^{-1}$ of 002$_{PV}$ and decrease in (integration width)$^{-1}$ of 008$_{BM}$ is reflecting progressive oxidation in SrCoO$_x$, since there is no systematic change in the (integration width)$^{-1}$, the change in the integration width of 008$_{BM}$ or 002$_{PV}$ Bragg diffraction peak seems not to reflect the grain size. It should be noted that the lateral grain size of both BM-SrCoO$_{2.5}$ (14.5 ±0.4 nm) and PV-SrCoO$_3$ (10.4 ±1.3 nm) is obviously smaller than the film thickness (40 nm). Although these XRD results roughly suggest that columnar oxidation of SrCoO$_{2.5}$ occurred, they are not enough to visualize the electrochemical oxidation of SrCoO$_x$.

In order to visualize the electrochemical oxidation of SrCoO$_x$, we measured the electrical resistivity ($\rho$) [**Fig. 4(a)** and **Supplementary Fig. S5**] and thermopower ($S$) [**Fig. 4(b)**] of the SrCoO$_x$ as a function of the conversion rate. The as grown sample showed insulating $\rho - T$ behaviour, which is similar with that of SrCoO$_{2.5}$ reported by Lu *et al.*[5] With increasing the oxidation state, the $\rho - T$ approached to that of SrCoO$_{3-\delta}$. From these results, we further confirmed that $x$ in the SrCoO$_x$ films could be modulated from 2.5 to 3 by the electrochemical oxidation. The as-grown sample (0 %) showed the $\rho$ of 4.0 $\Omega$ cm and the $S$ of +70 μV K$^{-1}$, whereas 100 % converted sample showed $\rho$ = 0.7 m$\Omega$ cm and $S$ = −0.5 μV K$^{-1}$ at room temperature. Both $\rho$ and $S$ drastically decreased with increasing conversion rate. Note that the sheet resistance of GDC/YSZ substrate is extremely high (~10$^{16}$ $\Omega$, 0.5 mm) as compared to that of SrCoO2.5 (<10$^7$ $\Omega$), the SrCoO$_{2.5}$ film (~40 nm) solely contributes the observable $S$.

We then calculated the $\rho$ and $S$ as a function of the conversion rate by assuming that the electric circuit is a parallel circuit composed of the insulating SrCoO$_{2.5}$ region and the conducting SrCoO$_{3-\delta}$ region as schematically shown in **Fig. 4(c)**.

$$(\rho_{para})^{-1} = \frac{(\rho_{SrCoO2.5})^{-1} \cdot (100-y) + (\rho_{SrCoO3})^{-1} \cdot y}{100}$$

and

$$S_{para} = \frac{S_{SrCoO2.5} \cdot (\rho_{SrCoO2.5})^{-1} \cdot \frac{100-y}{100} + S_{SrCoO3} \cdot (\rho_{SrCoO3})^{-1} \cdot \frac{y}{100}}{(\rho_{SrCoO2.5})^{-1} \cdot \frac{100-y}{100} + (\rho_{SrCoO3})^{-1} \cdot \frac{y}{100}} \quad [17, 26]$$

, where $y$ is the conversion rate. The dotted lines in Figs. 4(a) and 4(b) were calculated by assuming the parallel electrical circuit of the insulating SrCoO$_{2.5}$ region and the conducting SrCoO$_{3-\delta}$ region. Although the dotted line reproduces well both $\rho$ and $S$ when the conversion rate is above 60 %, the assumed line does not reproduce the observed $\rho$ and $S$ when the conversion rate is less than 60 %. The system is best described by the isolated conducting

SrCoO$_{3-\delta}$ columns surrounded by the insulating SrCoO$_{2.5}$ matrix. Thus, contribution of series circuit of the insulating SrCoO$_{2.5}$ region and the conducting SrCoO$_{3-\delta}$ region along the current direction is not negligible.

Then, we calculated the $\rho$ and $S$ as a function of the conversion rate by assuming the columnar oxidation occurred as schematically show in **Fig. 4(d)**. According to the data fitting using the percolation model [27] of SrCoO$_{3-\delta}$ region, we reproduced the observed $\rho$ and $S$ very well.

$$\rho_{percolation} = \rho_{SrCoO2.5} \cdot \left(1 - \frac{\phi}{\phi_c}\right) \qquad (\phi_c < 0.25)$$

$$\rho_{percolation} = \frac{1}{\rho_{SrCoO3-\delta}} \cdot (\phi - \phi_c) \qquad (\phi_c > 0.25)$$

$$S_{percolation} = S_{SrCoO2.5} \cdot \left(1 - \frac{\phi}{\phi_c}\right) \qquad (\phi_c < 0.25)$$

$$S_{percolation} = S_{SrCoO3-\delta} \cdot \left(\frac{\phi}{\phi_c}\right) \qquad (\phi_c > 0.25)$$

where $\phi$ is the conversion rate and $\phi_c$ is the percolation threshold. From these results, we concluded that the columnar oxidation firstly occurred along with the surface normal and then spread in the perpendicular to the normal direction.

Then, we directly observed the conducting region using the conductive AFM to confirm our conclusion described above. **Figure 5** shows the macroscopic mapping of the insulating SrCoO$_{2.5}$ region and the conducting SrCoO$_{3-\delta}$ region in the partially oxidized SrCoO$_x$ films. Although their topographic AFM images (2 μm × 2 μm) did not show any difference each other, the difference in the electric current mappings was clearly visualized [**Figs. 5(e) – 5(h)**]. Connection of the conducting region in the images (g) and (h) is clearly visible whereas that in the images (e) and (f) is not seen. Note that lateral dimension of each highly conducting area can stretch up to ~200 nm. So, it is likely that along the depth direction the sample is fully oxidized, if we assume velocity of oxidation is uniform. According to the percolation theory, parallel electrical circuit composed of insulating $SrCoO_{2.5}$ region and conducting $SrCoO_{3-\delta}$ region can be assumed in the case of images (g) and (h) with uniform and high current distribution on the plane. However both parallel and series circuit composed of insulating $SrCoO_{2.5}$ region and conducting $SrCoO_{3-\delta}$ region contribute in the case of the images with dot-like current distribution from (e) and (f), which are below the percolation threshold. These observations are fully consistent with the resistivity and thermopower values.

From these results, we successfully visualized the electrochemical oxidation of $SrCoO_{2.5}$ film into $SrCoO_{3-\delta}$ film. First, the columnar oxidation occurred perpendicular to the film surface and then spread in the lateral direction. This macroscopic image of the electrochemical oxidation would be useful to develop a functional device utilizing the electrochemical redox reaction of $SrCoO_x$ in the future.

In summary, we investigated the electrochemical oxidation reaction of $SrCoO_x$ so-called 'oxygen sponge'. Thermopower analysis and conductive AFM analyses were performed to clarify the electrochemical oxidation macroscopically. $SrCoO_x$ epitaxial films with various oxidation states were prepared by the electrochemical oxidation of $SrCoO_{2.5}$ into $SrCoO_3$. The XRD results suggested that columnar oxidation occurred. The resistivity and thermopower data can be fitted by assuming the percolation of conducting $SrCoO_3$ columns. When the column density of $SrCoO_3$ exceeds 25 %, the resistivity and thermopower dramatically decreased. The flown current and the conducting area in the conductive AFM image increased with increasing oxidation state. The columnar oxidation firstly occurred along with the surface normal and then spread in the normal direction. This macroscopic image of the electrochemical oxidation would be useful to develop a functional device utilizing the electrochemical redox reaction of SrCoO$_x$.

**Experimental Section**

*Pulsed laser deposition of the films*: The films were heteroepitaxially grown on (001) Yttria−Stabilized Zirconia (YSZ) single crystal substrate by pulsed laser deposition (PLD). During the film deposition, the substrate temperature was kept at 750 °C and the oxygen pressure was kept at 10 Pa. KrF excimer laser pulses (~2 J cm$^{-2}$ pulse$^{-1}$, 10 Hz) were irradiated onto the ceramic target. First, 6-nm-thick 8 mol% Gd-doped CeO$_2$ (GDC) was grown followed by the growth of ~40-nm-thick SrCoO$_{2.5}$.[22-24] In order to check the quality of the resultant film, we observed the reflection high energy electron diffraction patterns of the resultant films before air exposure.

*Device fabrication and electrochemical oxidation*: 55-nm-thick porous Pt film (relative density ~53%) with comb shaped pattern was deposited on the back side of the YSZ substrate by the d.c. sputtering (IB-3, Eiko Co.) at room temperature. As the counter electrode, we mechanically attached an Al foil as schematically shown in **Fig. 1b**. In order to oxidize the SrCoO$_{2.5}$ into SrCoO$_{3-\delta}$, the sample was heated at 300 °C in air and the oxidative current was applied under applying −5 V or −10 V. The current density was up to −20 μA cm$^{-2}$. The electrochemical reaction is denoted by the following formula.

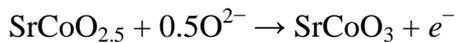

$$\text{SrCoO}_{2.5} + 0.5\text{O}^{2-} \rightarrow \text{SrCoO}_3 + e^-$$

Thus, the electron density of $1.61 \times 10^{22}$ cm$^{-3}$ is needed to fully oxidize SrCoO$_{2.5}$ into SrCoO$_3$. We calculated the required quantity of electricity per unit area as $1.61 \times 10^{22} \times$ (film thickness of SrCoO$_{2.5}$) $\times 1.602 \times 10^{-19}$ = 9.5 mC cm$^{-2}$. We modulated the oxidation state by changing the electron density as 0.44 mC cm$^{-2}$ (4.6 %) for sample A, 0.88 mC cm$^{-2}$ (9.3 %) for sample

B, 2.1 mC cm$^{-2}$ (22 %) for sample C, 4.0 mC cm$^{-2}$ (42 %) for sample D, and 12 mC cm$^{-2}$ (126 %) for sample E, respectively.

*Crystallographic analyses*: The crystalline phase, orientation, lattice parameters, and thickness of the films were analyzed by high resolution X-ray diffraction (XRD, Cu Kα$_1$, ATX-G, Rigaku Co.). Out-of-plane Bragg diffraction patterns and the rocking curves were measured at room temperature. The X-ray reciprocal space mappings (RSMs) were also recorded to clarify the change of the SrCoO$_x$ lattice. X-ray reflection patterns were measured to evaluate the density and the thickness. An atomic force microscopy (AFM, Nanocute, Hitachi Hi-Tech Sci. Co.) was used to observe the surface microstructure of the films.

*Resistivity and thermopower measurements*: Resistivity of the resultant films was measured by dc four-probe method with van der Pauw electrode configuration. In-Ga alloy was used as the contact electrodes. Thermopower of the resultant films was measured by standard steady state method. The film sample was placed on the gap (~5 mm) between two Peltier devices. By applying the forward/reverse current to each Peltier devices, temperature difference was generated in the sample. We measured the temperature difference ($\Delta T$) and the thermos-electromotive force ($\Delta V$) simultaneously at room temperature. The thermopower was calculated as the linear slope of $\Delta T - \Delta V$ plot.

*Macroscopic mapping of insulating region and conducting region*: Conducting region of the film surface was visualized using conductive AFM (MFP-3D Origin$^{TM}$, Oxford Instruments) by applying −2 V against the sample surface using Pt coated cantilever. The scanned area was 5 μm × 5 μm (512 lines, 1 Hz). The measurement was performed in air at room temperature.

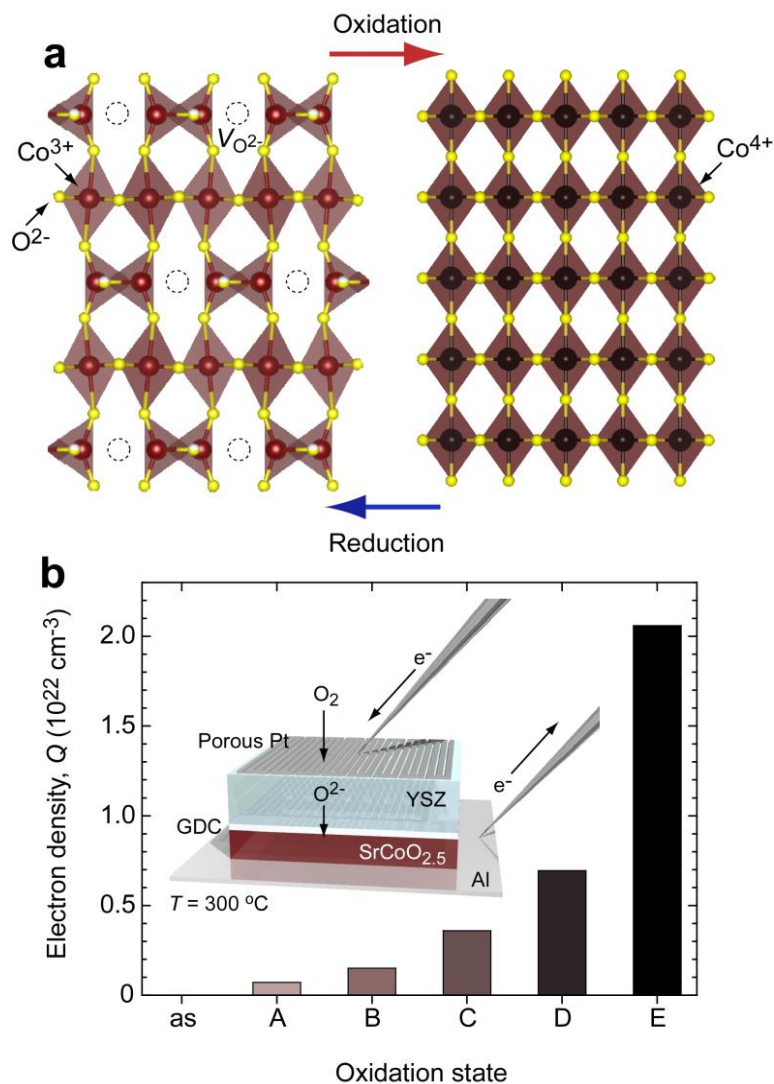

**Figure 1.** Electrochemical reaction of SrCoO$_x$ oxygen sponge. (a) Schematic crystal structure of brownmillerate SrCoO$_{2.5}$ and perovskite SrCoO$_{3-\delta}$. Although there are oxygen vacancies ($V_{O2-}$) in SrCoO$_{2.5}$ crystal, the vacancies are occupied with oxygen after the electrochemical oxidation (SrCoO$_{3-\delta}$). Valence states of cobalt ion change from 3+ into 4+ by the oxidation. This electrochemical reaction is reversible. (b) SrCoO$_x$ samples with different oxidation states. Oxidation current with different electron density was applied to the multilayer structure composed of Pt/YSZ/GDC/SrCoO$_{2.5}$/Al at 300 °C in air as shown in the inset.

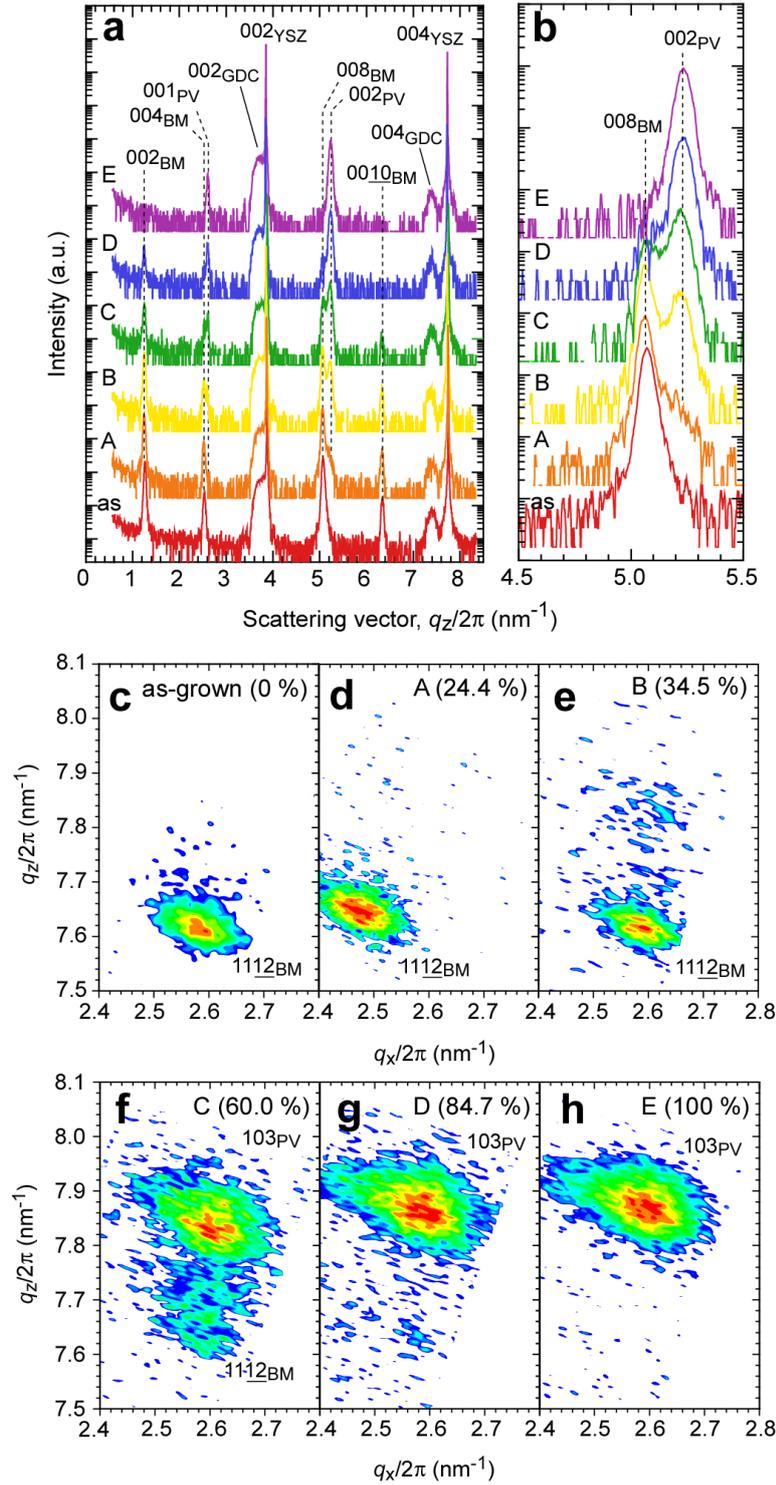

**Figure 2.** X-ray diffraction analyses of SrCoO$_x$ films with various oxidation states. (a) Out-of-plane XRD patterns. Intense diffraction peaks of 00$l_{BM}$ or 00$l_{PV}$ are seen together with 00$l_{GDC}$/00$l_{YSZ}$, indicating strong c-axis orientation of the films. (b) Magnified XRD pattern. With increasing the oxidation state, the peak position changed from 008$_{BM}$ to 002$_{PV}$. (c)−(h) RSMs around 11$\underline{12}_{BM}$ or 103$_{PV}$. Although the diffraction spot was 11$\underline{12}_{BM}$ when the oxidation state is lower than 34.5 %, it changed into 103$_{PV}$ when the oxidation state was greater than 60 %.

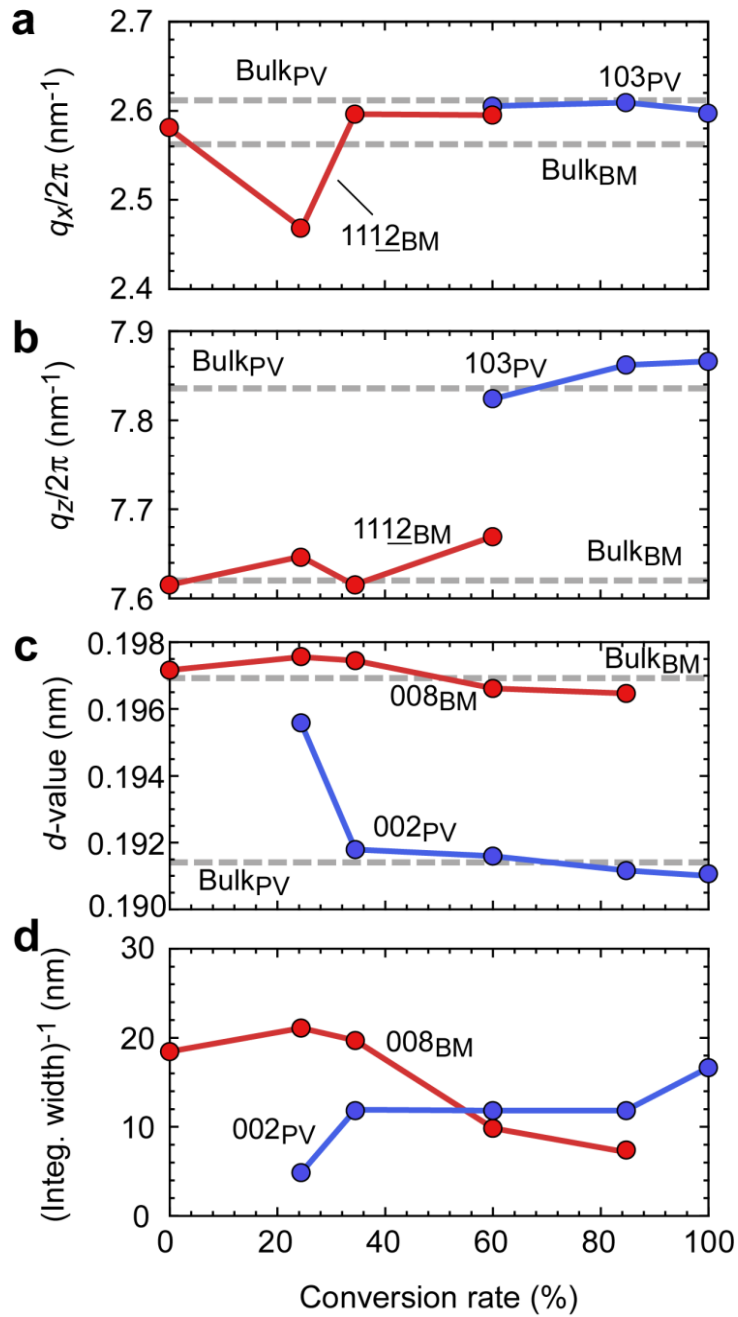

**Figure 3.** Change in the crystallographic size of the SrCoO$_x$ film as a function of the conversion rate. (a) Scattering vector in the in-plane direction ($q_x/2\pi$) and (b) scattering vector in the out-of-plane direction ($q_z/2\pi$) of 111$\underline{2}$$_{BM}$ and 103$_{PV}$. (c) $d$-value and (d) inverse of the integral width of 008$_{BM}$ and 002$_{PV}$. Crystal phase changed from BM to PV around 40 − 60 %. The conversion rate dependences of the inverse of the integral width suggest that the columnar oxidation of SrCoO$_x$ occurred.

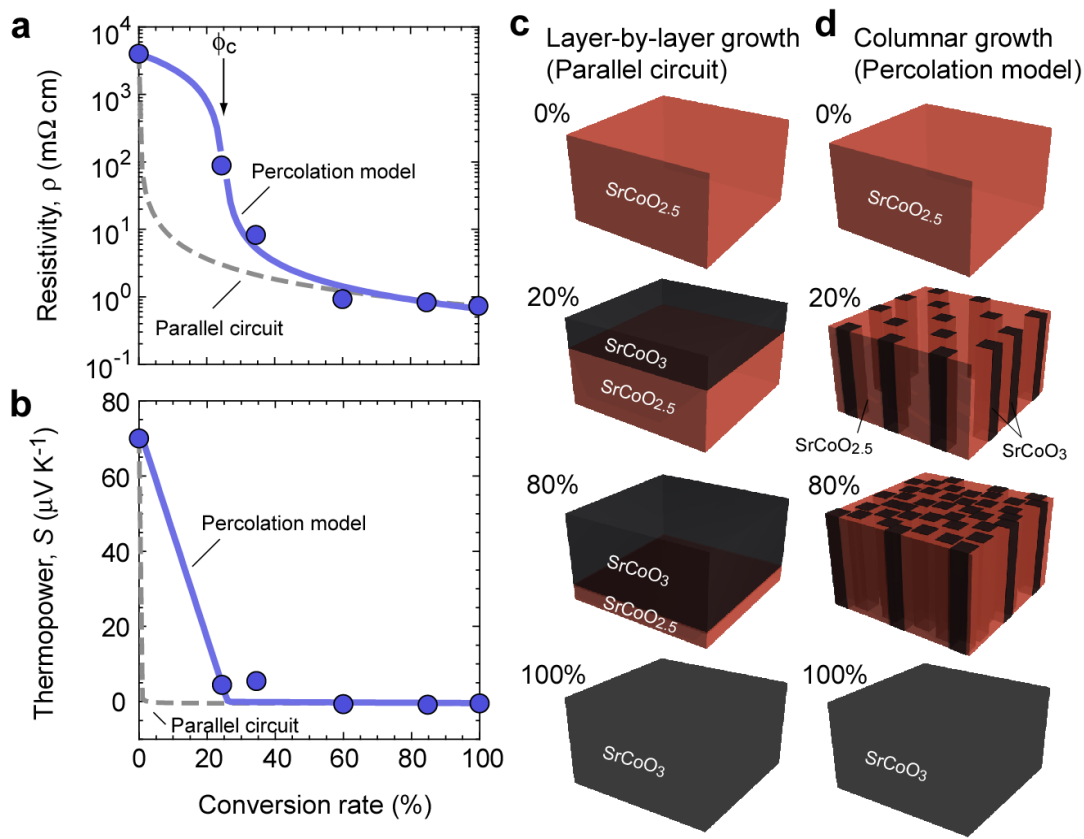

**Figure 4.** Electron transport properties of the SrCoO$_x$ as a function of the conversion rate at room temperature. (a) Resistivity and (b) thermopower. The as grown sample (0 %) showed the resistivity of 4.0 Ω cm and the thermopower of +70 μV K$^{-1}$, whereas 100 % converted sample showed 0.7 mΩ cm and −0.5 μV K$^{-1}$. The dotted lines were calculated by assuming (c) the parallel electrical circuit of insulating SrCoO$_{2.5}$ region and conducting SrCoO$_{3-\delta}$ region. The solid lines were calculated by assuming (d) the percolation model (columnar growth) of conducting SrCoO$_{3-\delta}$ region. The percolation model reproduces the observed values well, indicating columnar growth of SrCoO$_{3-\delta}$ occurred.

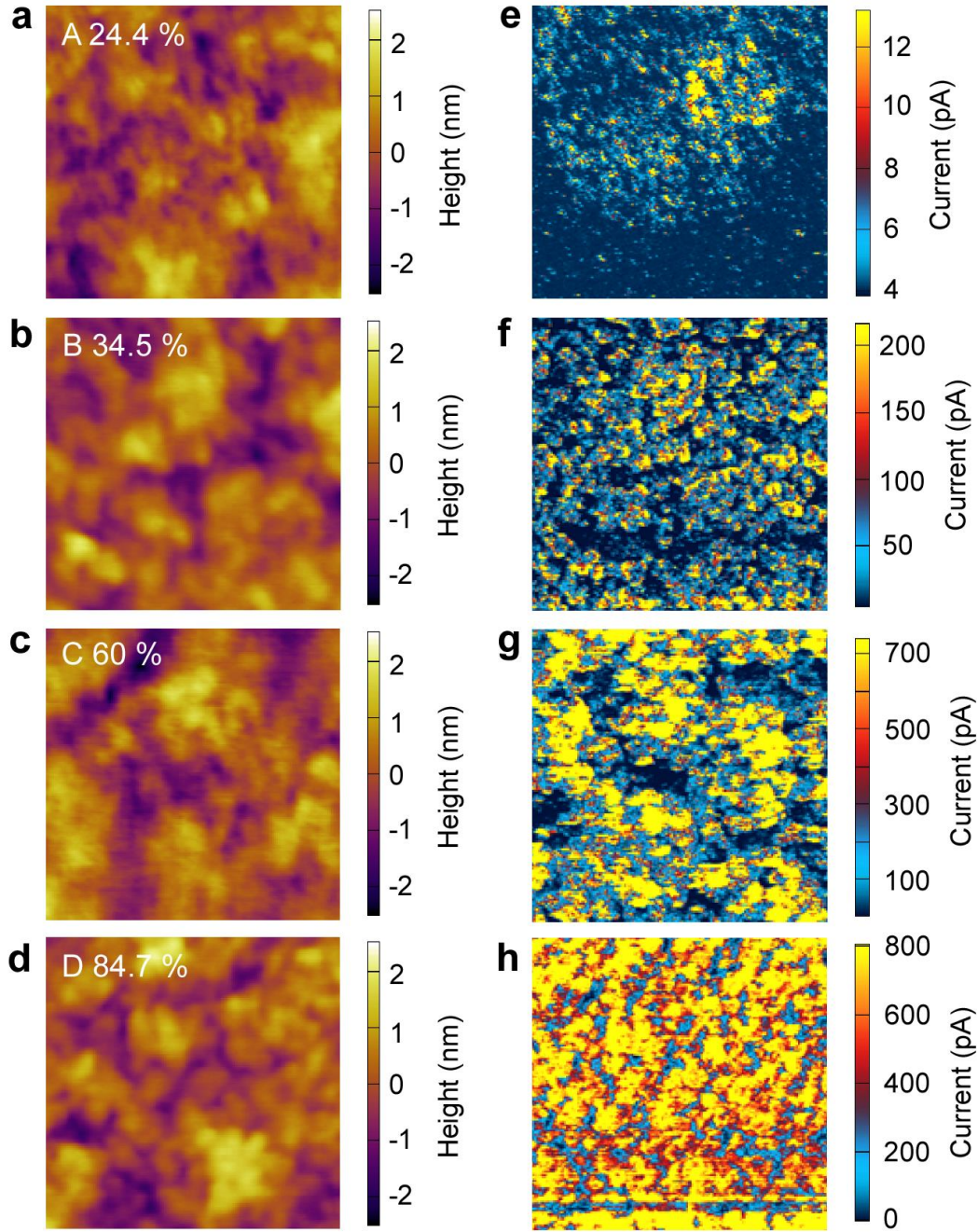

**Figure 5.** Macroscopic mapping of insulating $SrCoO_{2.5}$ region and conducting $SrCoO_{3-\delta}$ region. Topographic AFM images (2 μm × 2 μm) of (a) sample A (24.4 %), (b) sample B (34.5 %), (c) sample C (60 %) and (d) sample D (84.7 %). (e − h) Electric current mapping of the samples corresponding to the topographic AFM images of (a) − (d). The currents were taken under applying −2 V against the sample surface. Connection of the conducting region in the images (g) and (h) is clearly visible whereas that in the images (e) and (f) is not seen. According to the percolation theory, parallel electrical circuit composed of insulating $SrCoO_{2.5}$ region and conducting $SrCoO_{3-\delta}$ region can be assumed in the case of images (g) and (h). However both parallel and series circuit composed of insulating $SrCoO_{2.5}$ region and conducting $SrCoO_{3-\delta}$ region contribute in the case of the image (e), which are below the percolation threshold ($\phi_c$ = 25 %).

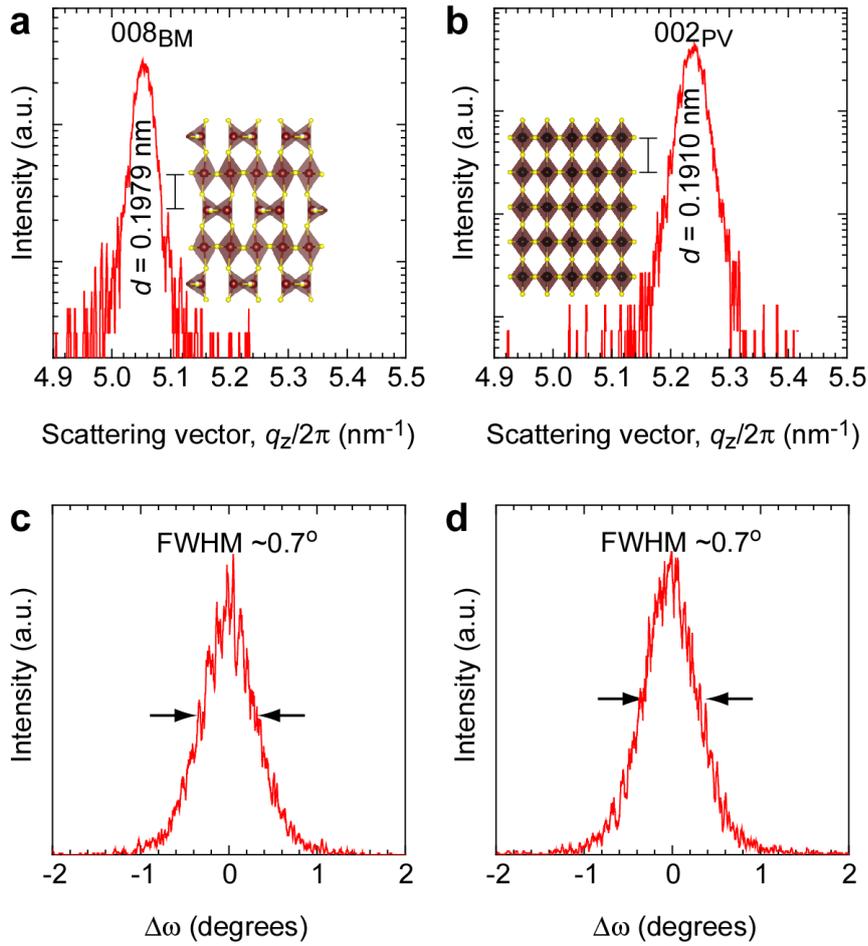

**Figure S1.** Out-of-plane lattice parameter and tilting of SrCoO$_x$ films at room temperature. (a)(b) Out-of-plane Bragg diffraction patterns of (a) as-grown ($x$ = 2.5, 008$_{BM}$) and (b) electrochemically oxidized ($x$ = 3−δ, 002$_{PV}$) SrCoO$_x$ films. The *d*-spacing of 008$_{BM}$ is 0.1979 nm whereas that of 002$_{PV}$ is 0.1910 nm, ~3.2 % shorter than 008$_{BM}$. (c)(d) Out-of-plane X-ray rocking curves of (c) 008$_{BM}$ for the as-grown and (d) 002$_{PV}$ for the electrochemically oxidized SrCoO$_x$ films. Average tilting (≡ full width at half maximum, FWHM) of both SrCoO$_x$ films is ~0.7 °, which does not show any difference between as-grown and oxidized.

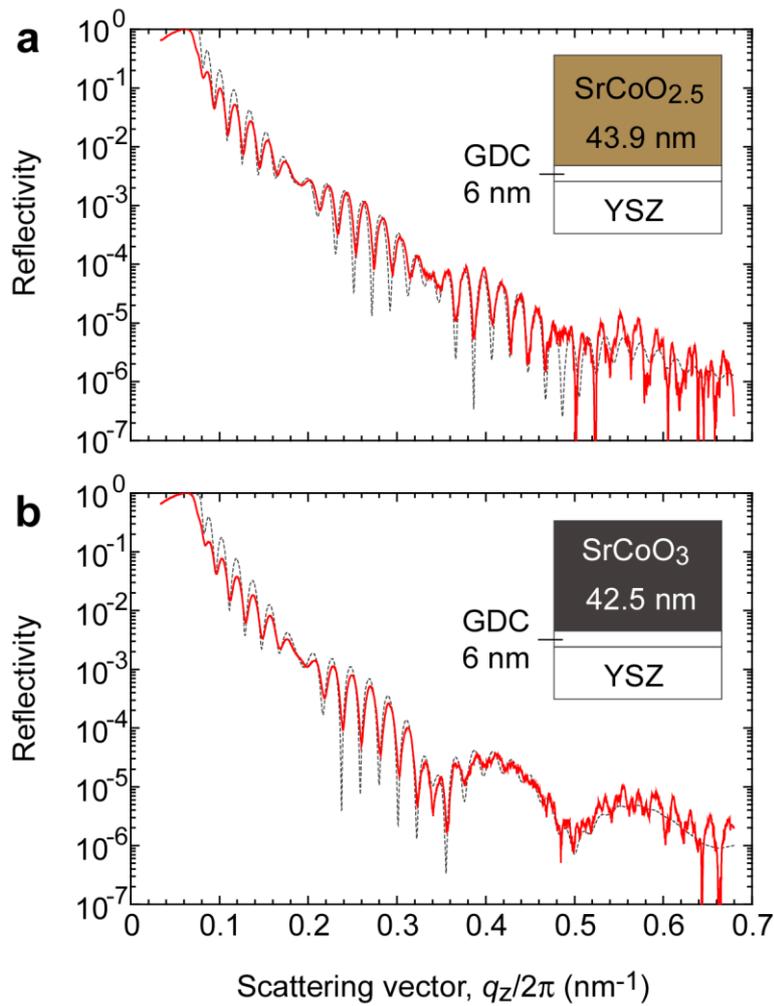

**Figure S2.** Characterization of the SrCoO$_x$/GDC bilayer structure on YSZ substrate. X-ray reflectivity of (a) as grown SrCoO$_{2.5}$/GDC and (b) oxidized SrCoO$_{3-\delta}$/GDC (red line: observed, gray dotted line: simulated). As schematically shown in the inset, the thickness of the SrCoO$_{2.5}$ film was 43.9 nm whereas that of the SrCoO$_{3-\delta}$ was 42.5 nm. The film thickness decreased ~3.2 % due to decreasing of the d-spacing, indicating that the SrCoO$_{2.5}$ was fully oxidized into SrCoO$_3$. Thickness of the GDC film did not change after the oxidation.

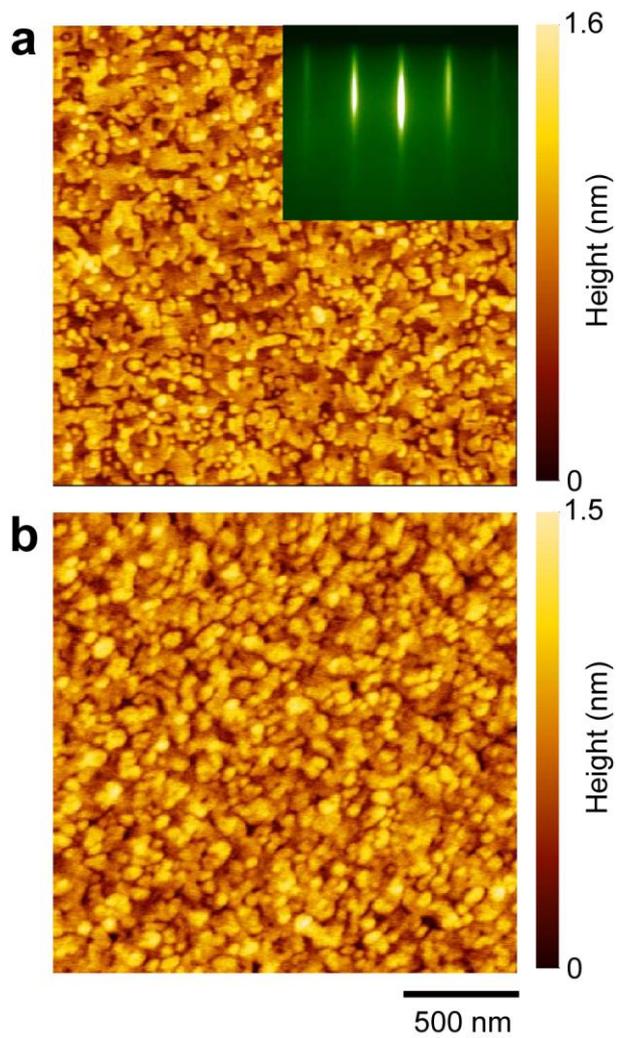

**Figure S3.** Surface morphology of the SrCoO$_x$ films. Topographic AFM images of (a) as grown SrCoO$_{2.5}$ and (b) oxidized SrCoO$_{3-\delta}$ films (2 μm × 2 μm). Less than one hundred nanometer sized grain structure is seen in both images. The surface morphologies are basically same. Bright streaks in the RHEED pattern indicate that the SrCoO$_{2.5}$ was heteroepitaxially grown on the substrate with smooth surface.

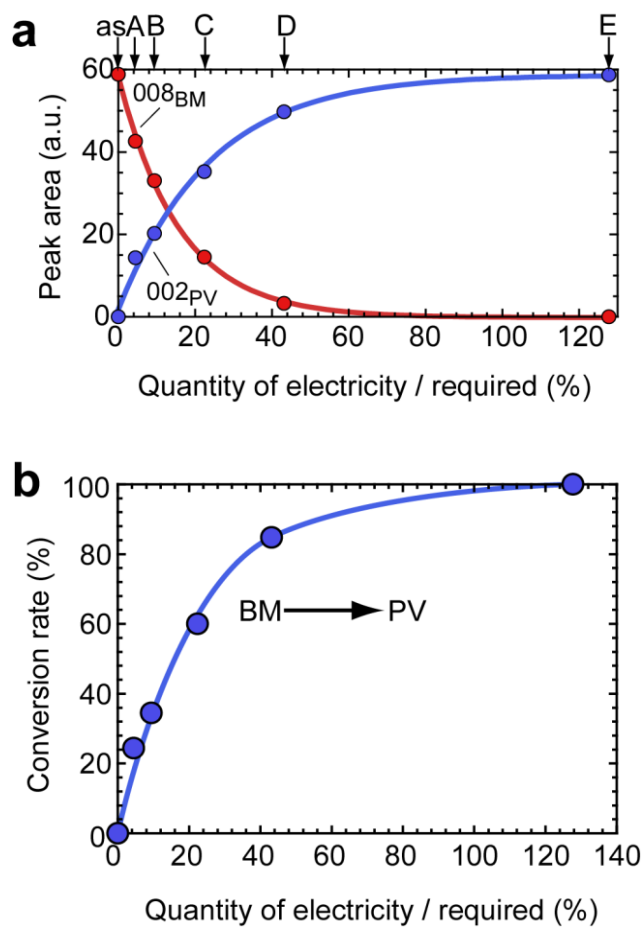

**Figure S4.** Calculation of the conversion rate from BM-SrCoO$_{2.5}$ into PV-SrCoO$_3$. (a) Changes in the diffraction peak area of (red) 008$_{BM}$-SrCoO$_{2.5}$ and (blue) 002$_{PV}$-SrCoO$_{3-\delta}$ as a function of the quantity of electricity over required electricity. The required electricity is 9.6 mC. With increasing the electricity, the peak area of 002PV increases rapidly and saturates whereas that of 008BM decreases rapidly and approaches to zero. (b) Conversion rate from BM-SrCoO$_{2.5}$ into PV-SrCoO$_{3-\delta}$. We assumed that **sample E** was fully oxidized. The conversion rate was calculated as (peak area of 002$_{PV}$)/(peak area of 002$_{PV}$ for **sample E**).

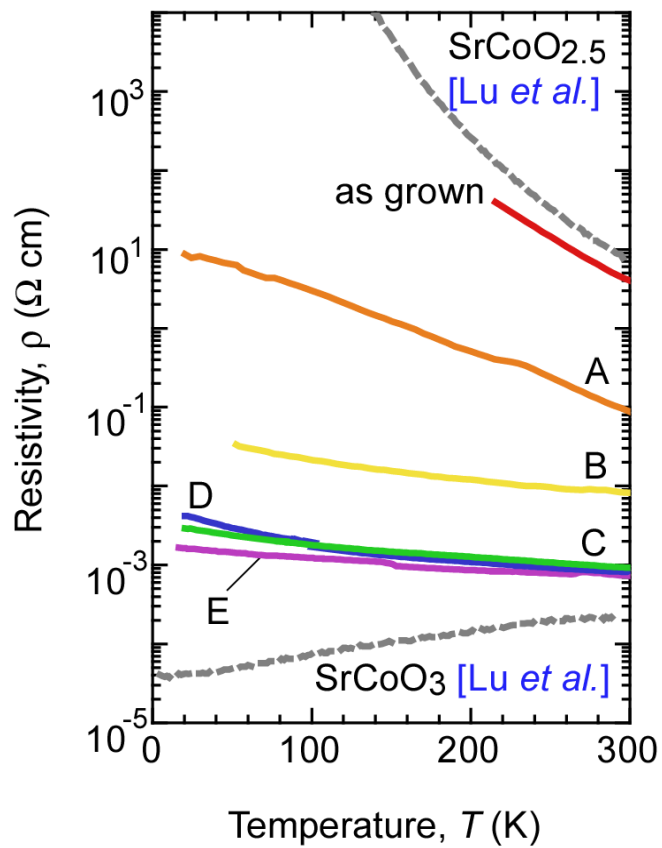

**Figure S5.** Temperature dependence of the resistivity of the $SrCoO_x$ films. The as grown sample shows insulating $\rho-T$ behaviour, which is similar with that of $SrCoO_{2.5}$ reported by Lu et al.[1] With increasing the oxidation state, the $\rho-T$ apploaches to that of $SrCoO_3$ reported by Lu et al.